\documentstyle[12pt,a41,epsfig,portland,rotating,amstex,cite]{article}

\newcommand{\beq}{\begin{equation}}
\newcommand{\eeq}{\end{equation}}
\newcommand{\bea}{\begin{eqnarray}}
\newcommand{\eea}{\end{eqnarray}}

\newcommand{\lsim}{\raisebox{-0.07cm}{$\, \stackrel{<}{{\scriptstyle
\sim}}\, $}}
\newcommand{\gsim}{\raisebox{-0.07cm}{$\, \stackrel{>}{{\scriptstyle
\sim}}\, $}}

\newcommand\GeV{\,\mbox{GeV}}

\newcounter{lin}

\graphicspath{{./},
	      {./figures/}}

\begin{document}
\begin{titlepage}

\begin{flushleft}
DESY 05--201 \hfill {\tt hep-ph/0510212} \\
SFB-CPP-05-67 \\
\end{flushleft}

\vspace{3cm}
\noindent
\begin{center}
{\LARGE\bf Status of Polarized and Unpolarized}

\vspace{2mm}
{\LARGE\bf Deep Inelastic Scattering\footnote{Invited talk, DIS 2005, 
Madison, WI; Duality05, LNFN, Frascati, Italy.}}
\end{center}
\begin{center}

\vspace{2.0cm}
{\large Johannes Bl\"umlein }

\vspace{1.5cm}
{\it 
Deutsches Elektronen Synchrotron, DESY}\\

\vspace{3mm}
{\it  Platanenallee 6, D--15738 Zeuthen, Germany}\\

\vspace{3cm}
\end{center}
\begin{abstract}
\noindent
The current status of deep inelastic scattering is briefly reviewed.
We discuss future theoretical developments desired and measurements 
needed to further complete our understanding of the picture of nucleons 
at short distances. 
\end{abstract}

\end{titlepage}

\newpage
\sloppy
\section{Introduction}

\vspace{1mm}\noindent
The discovery of the partonic substructure of nucleons by the SLAC--MIT 
experiments \cite{SLACMIT} 35 years ago marks the beginning of the 
investigation of the nucleon's short distance structure. During the 
subsequent decades numerous $e^{\pm} N$, $\mu^{\pm} N$ and $\nu 
(\overline{\nu}) N$--experiments were performed at SLAC, FNAL, CERN, DESY 
and JLAB both for unpolarized and polarized targets to refine our 
understanding of nucleons in wider and wider kinematic domains and at 
higher luminosities which allowed rather precise measurements.  

Along with this, the theoretical understanding deepened applying Quantum 
Chromodynamics (QCD) perturbativly to higher orders and investigating 
some of the related operator matrix elements with non--perturbative 
methods
in the framework of Lattice QCD over the last decades. Deep inelastic 
scattering data do allow for QCD tests at the 1\% level \cite{MK1P} at 
present, which requires $O(\alpha_s^3)$ accuracy for the perturbative 
calculations.

In the following I give a brief survey on the present status of deeply 
inelastic scattering (DIS) and discuss the current challenges for theory 
and experiment in this field.
\section{Theory}
\label{sec:theory}

\vspace{1mm}\noindent
The theory of deeply inelastic scattering has a history of about 40 years.
Beginning with the early work on the light cone expansion \cite{LCE}  and 
the 
parton model \cite{PM} a conclusive picture of the twist--2 contributions
\cite{TW} arose in complementary languages. With the advent of QCD 
\cite{QCD} and finding asymptotic freedom \cite{ASFR} the scaling 
violations of nucleon structure functions were studied systematically.
The leading order (LO) results for the anomalous dimensions (1973) 
\cite{PLO}
were followed by the LO coefficient functions, NLO anomalous dimensions and 
coefficient functions~\cite{NLO}, see Figure~1, until after about 20 years 
the 3--loop anomalous dimensions and coefficient functions could be 
calculated recently \cite{NNLO}. These calculations are required to match 
the current experimental accuracies, in particular to extract the QCD 
parameter $\Lambda_{\rm QCD}$ with a theoretical error below the 
experimental accuracy. Similar timescales 
were necessary to reach the 4--loop level for the QCD $\beta$-function.  
The step from NLO to NNLO expressions required a significant change in 
technology and intense use of efficient Computer Algebra programs like 
{\tt FORM} \cite{FORM} due to the proliferation of terms emerging. The 
calculus of harmonic sums and associated functions \cite{HSUM} was both 
helpful to design a uniform 
language for higher order calculations, led to a systematic approach, and 
gained deeper insight into what eventually is  really behind intermediate 
large expressions generated by Feynman diagrams. Still further progress 
has to 
be made in the future. QCD perturbation theory took an enormous 
development during the last three decades transforming our understanding 
from an initially {\sf qualitative} one to highly a {\sf quantitative} 
level. Physics, as a quantitative science, knows no other ways but 
precision calculations to put theoretical ideas and theories to an 
utmost check. In this way, Quantum Chromodynamics became a well tested, 
established physical theory, which of course is a process to be 
continued steadily. The mathematical methods being developed in 
course to 
perform this task have a deep beauty and give us insight into quantum 
field theory on a meta-level. Their uniform applicability has 
furthermore led to a quick spread into a series of neighboring fields, as 
electro--weak theory and string theory, 
and became, only a few years after their development, a common tool of the 
community.

\vspace{1mm}
With the advent of HERA it became possible to probe the small--$x$ region. 
Much work was devoted to 
resummations in this particular region. As pioneered by Lipatov 
and 
collaborators \cite{BFKL} both the leading contributions to the splitting 
and coefficient functions can be derived by arguments of scale invariance.

\vspace*{2mm}
\begin{center}
\mbox{\epsfig{file=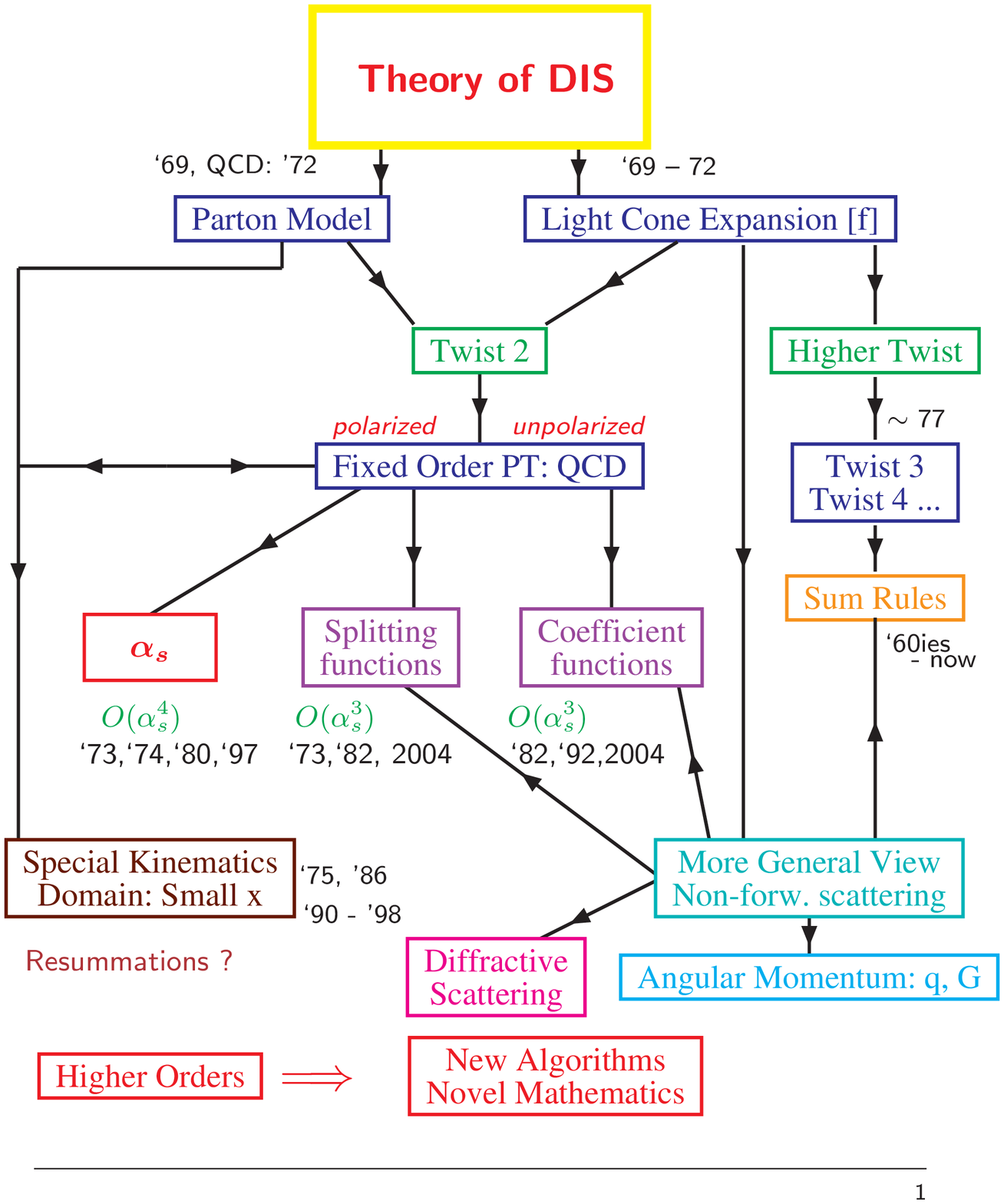,width=\linewidth}}

\vspace{1mm}
\noindent     
\small   
\end{center}
{\sf
Figure~1:~
The development of the theory of deeply inelastic scattering. The years 
refer to the respective calculations.}
\normalsize

Resummed NLx corrections were calculated in \cite{NLx}. All these results 
are of very importance as all--order predictions for the leading and 
next-to-leading term for splitting functions and Wilson coefficients. 
The comparison of these predictions with the corresponding results 
obtained in complete fixed order calculations showed agreement. 
Furthermore one should stress, that the LO resummations \cite{BFKL} refer 
to {\sf scheme-invariant}, i.e. physical quantities, and do thus predict 
as well corresponding matching conditions between splitting and 
coefficient functions.

Unfortunately these resummations turn out to be not dominant in the small 
$x$ region for the description of structure functions, since the 
respective kernels have to be convoluted with parton 
densities, which are strongly rising towards small values of $x$ and 
formally sub-leading terms contribute at the same strengths  \cite{BV}.

For polarized deep--inelastic scattering the anomalous dimensions and 
coefficient functions are known to NLO at present. Although the 
statistical and systematic errors for the polarized parton densities is 
still large, the NNLO improvement would be desirable to further minimize 
the factorization and renormalization scale uncertainties~\cite{OSAKA}.
  
Polarized deep inelastic scattering offers access to twist--3 operator 
matrix elements and predictions for their scale dependence in QCD. At 
present, the radiative corrections for these terms are worked out in 
one-loop order. The understanding of QCD higher twist contributions beyond 
twist 3 both for unpolarized and polarized deep inelastic scattering is 
still in its infancy and will require more work in the future 
having more precise data available. Since higher twist anomalous 
dimensions and coefficient functions refer to more than one ratio of 
scales but structure functions contain one scale, $x_{\rm Bj}$, only, it 
is required in general to measure the corresponding operator matrix 
elements on the lattice at least for the lowest moments.

The light--cone expansion as used in deep inelastic scattering can be 
generalized to a series of other processes. This more general view 
concerns non--forward scattering at large space--like virtualities 
\cite{NONF}. In this way one may access the angular momentum of partons 
\cite{JI},
which is important for the understanding of the spin--structure of 
nucleons. Moreover, the framework provides several projections on various 
inclusive quantities of interest. 
During the last decade a lot of progress has been made in 
this field calculating the corresponding LO and NLO anomalous dimensions 
and Wilson coefficients both for the unpolarized and polarized case and 
understanding conformal symmetry and its breaking for this process in 
QCD. A related picture was developed also to describe diffractive 
$ep$--scattering \cite{DIFFR}, which yields a proper description of this 
process using the notion of an {\sf observed} rapidity gap only, without 
referring to the concept of a pomeron.

\section{Experiment}
\label{sec:experiment}

\vspace{1mm}\noindent
Deep inelastic scattering has been probed by now in a wide kinematic 
range~: $10^{-5} < x < 0.8,~~4 < Q^2 \lsim  50.000 \GeV^2$. Figure~2 
gives 
an overview on different experiments and facilities showing also the 
luminosities reached or planned. The proton structure function 
$F_2^p(x,Q^2)$ is a well measured quantity in all this range. Both to 
perform flavor separation and QCD tests, it is highly desirable to know 
the 
neutron structure function $F_2^n(x,Q^2)$ \cite{BKNR} at comparable 
accuracy in the 
same kinematic region. This has been the case for fixed target 
experiments. Both measurements allow to extract the $u_{\rm val}$ and 
$d_{\rm val}$ distributions at comparable precision not only in the 
valence region $x \gsim 0.3$ but also in the region below, supplementing 
the DIS data with Drell-Yan data on $\overline{d}(x) - \overline{u}(x)$. 
A non--singlet QCD analysis to $O(\alpha_s^3)$ was performed 
\cite{BBG}, widely free on assumptions on the gluon and sea-quark 
densities. The error of $\alpha_s$ is of $\sim 3\%$.

\begin{center}
\mbox{\epsfig{file=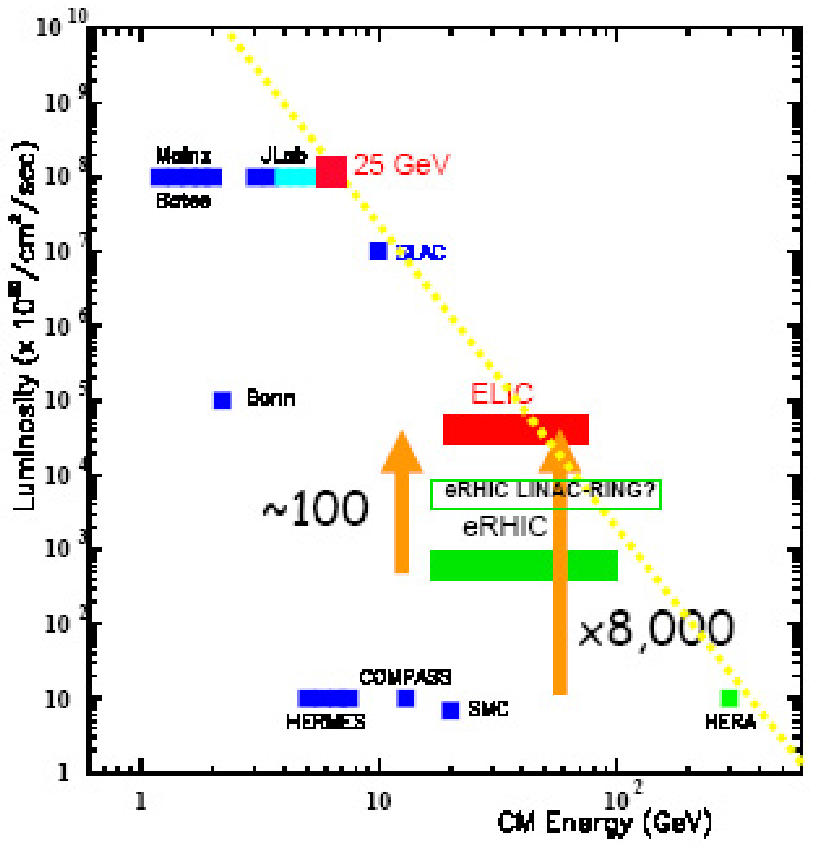,width=15cm}}

\vspace{2mm}
\noindent     
\small   
{\sf
Figure~2:~
DIS $lN$ scattering experiments: Luminosity vs cms Energy [curtesy R. 
Ent]. 
}
\end{center}

\normalsize

\vspace{2mm}
The HERA experiments extended the kinematic region by two orders of 
magnitude both in $x$ and $Q^2$. From the high $Q^2$ large $x$ data   
the valence distributions will be measured in a yet widely unexplored 
region, 
in which potentially new physics may be found. These measurements can 
be compared to those at lower values of $Q^2$ by evolution. Statistically 
very precise measurements were performed in the medium and lower $Q^2$ 
and small $x$ region, which gives access to the charge--weighted 
sea--quark 
and gluon distributions. As the detailed knowledge of the gluon and
sea--quark distributions is instrumental to the future physics at LHC, 
HERA plays a 
key role in determining these quantities. In the case of $F_2(x,Q^2)$ the 
gluon distribution enters indirectly and determines the slope in 
$\ln(Q^2)$ of the structure function rather its value.  
From the measurement of both the slope and value of the structure function 
one may uniquely unfold the gluon and sea quark contribution without 
reference to a priori choices of shapes, cf.~\cite{JBAG}. Further 
important input for a measurement of the gluon distribution come from 
precise data on $F_2^{c\overline{c}}(x,Q^2)$ and $F_L(x,Q^2)$. For both 
these structure functions the gluon distribution enters linearly already 
in the lowest order. Combined non-singlet and singlet NNLO QCD analyses, 
partly including collider data, were performed in \cite{SA,MRST} measuring
$\alpha_s$ to a precision of $2-3 \%$ with central values in complete 
accordance with that of the non--singlet analysis \cite{BBG}.

The singlet quark distribution can well be extracted from $ep$--structure 
function measurements. On the other hand, the flavor structure of the sea 
quark distributions is hard to be resolved in neutral current 
interactions. Here future high luminosity neutrino experiments 
at high energy will  contribute. Drell--Yan data \cite{DY} provide 
information 
on the difference $\overline{d}(x) - \overline{u}(x)$. Still higher 
precision data is needed to resolve as well the $Q^2$ dependence. 
Information about the strange quark distributions $s(x),~\overline{s}(x)$ 
is currently gotten in a rather indirect way from the di--muon sample in 
DIS neutrino scattering. The high statistics measurements stem from iron 
targets and very little is known on the EMC--effect on strangeness in the 
lower $x$ region. The charm and beauty quark production in deep inelastic 
$ep$ scattering is well described by the heavy flavor Wilson coefficients
calculated to NLO \cite{HQNLO}. Very recently the NNLO corrections in the 
case of the longitudinal structure function $F_L^{Q\overline{Q}}$ for $Q^2 
\gg m^2$ were derived as the first result at $O(\alpha_s^3)$~\cite{FL3HQ}. 

The polarized deep inelastic parton densities are unfolded in QCD analyses
of the structure function $g_1^{eN}(x,Q^2)$, presently at NLO, 
\cite{POLDIS}. $\Delta u_v(x,Q^2)$ and $\Delta d_v(x,Q^2)$ are constrained 
best and an average statement can be made for the polarized sea under some 
assumptions as  $SU(3)_F$ symmetry or fixed ratios among some of the sea 
quark distributions. Flavor tagged measurements were performed 
\cite{FLTAG} to determine $\Delta q_i/q_i$ explicitely. These are first 
steps and higher luminosity measurements are required to reduce the errors 
further in the future. Constraints on the polarized gluon density are 
gotten through the QCD analysis and measuring open charm production. Yet 
the gluon density has a wide error band with  mainly positive central 
values. First experimental results were obtained for the transversity 
structure function $h_1(x,Q^2)$ \cite{TRANSV}. 
  
Important experimental tests concern the search for twist--3 contributions 
to polarized structure functions. For purely photonic interactions they 
are present in the structure function $g_2(x,Q^2)$ and, in the low 
$Q^2$ 
region, due to target mass effects, also in $g_1(x,Q^2)$~\cite{BT}. 
Similar 
to the Wandzura-Wilczek relation and other twist--2 relations in case of 
electro--weak interactions, the twist 3 contributions are related by 
integral relations, which can be tested in high luminosity experiments 
operating in the lower $Q^2$ region.

There is an ongoing programme to study deeply--virtual Compton scattering 
both at HERA and JLAB for a large set of observables. In course of these 
investigations one may hope to derive more information on the transverse 
sub--structure of nucleons and, potentially in the long term, information 
on parton angular momentum. 
\section{Future Avenues}
\label{sec:future}

\vspace{1mm}\noindent
Various important questions on the short--distance structure of nucleons
are yet open and require further experimentation and more theoretical 
work.
In the short run HERA will collect higher luminosity and measure
$F_2(x,Q^2), F_2^{c\overline{c}}(x,Q^2)$ with much higher precision. 
Different experiments will yield more detailed results on $g_2(x,Q^2)$ and 
the transversity distribution $h_1(x,Q^2)$. One of the central issues is 
to measure $F_L(x,Q^2)$ with high accuracy in different ranges of $Q^2$ 
and it would be essential to perform this measurement at HERA due to its 
unique kinematic domain. Much of our understanding of the gluon 
distribution depends on this measurement.

RHIC and LHC will lead to improved constraints on the gluon and sea 
quark distributions both for polarized and unpolarized nucleons. JLAB will 
contribute with high precision measurements in the large $x$ domain both 
for unpolarized and polarized nucleons, yet at low values of $Q^2$, which 
will increase with the advent of the higher energy option soon. These 
measurements supplement HERA's high precision measurements at small $x$. 
The 
possibility to investigate unpolarized and polarized deep inelastic 
scattering at the same experiments is crucial to minimize systematic 
errors. JLAB provides  ideal facilities to experimentally 
explore twist--3 effects in $g_2$ at high precision and to extract 
higher twist effects for unpolarized structure functions in unified  
analyses including data from large virtuality DIS at CERN and HERA.

For the time after HERA different $ep$ projects are 
discussed\footnote{Future high-luminosity muon-- and 
neutrino--factories will yield essential contributions to DIS, in 
particular concerning a detailed exploration of the sea-quark sector 
both for unpolarized and polarized nucleons. These projects are 
somewhat further ahead in time.}. 
Two of the projects are ERHIC and ELIC. The kinematic regions for 
both proposals is situated between the domain explored at CERN and 
HERA before, cf. Figure~2. While ERHIC reaches somewhat lower values of 
$x$, ELIC will 
have the higher luminosity, increasing that of HERA by a factor of 1000 to 
8000. As we saw before, various precision measurements are yet to be 
performed in this region. The programme will not concern inclusive 
quantities only but explore with sufficient luminosity also more rare 
channels, which are otherwise inaccessible but yield important theory 
tests. The high luminosity programme will answer many of the present 
open questions in the central kinematic region and yield  challenging 
non-trivial tests of Quantum Chromodynamics
both concerning perturbative and non--perturbative predictions.
On the theory side, perturbative higher order calculations will continue
at the level of $O(\alpha_s^3)$ corrections and heavy quark mass effects 
will be included. As much of the information during the next decade will 
come from proton--colliders, the respective processes have to be 
understood at higher precision. At the same time essential progress is 
expected in the systematic understanding of the moments of parton 
distributions on the lattice and for measuring the QCD scale $\Lambda_{\rm 
QCD}$. In this way, high luminosity experimental results, high order 
perturbative calculations, and highly advanced non-perturbative techniques
together will detail our understanding both of the strong force and
the nature of nucleons at short and longer distances. 

Particle physics always went along two avenues~: i) the search for new 
particles in annihilation processes at ever increasing energies; ii) the 
search for new sub--structures of matter in resolving shorter and 
shorter distances, the atomic nucleus and finally the nucleons.  
At present it seems that quarks are point--like 
particles. Since this may be  temporarily an impression, the search for 
their possible sub--structure has to be 
continued with suitable facilities in the future.
\newpage

\end{document}